\documentstyle[preprint,aps]{revtex}
\tightenlines
\begin{document}
\preprint{\today}
\title{Testing Mode-Coupling Theory for a Supercooled Binary
Lennard-Jones Mixture II: Intermediate Scattering Function and
Dynamic Susceptibility}
\author{Walter Kob\cite{wkob}}
\address{Institut f\"ur Physik, Johannes Gutenberg-Universit\"at,
Staudinger Weg 7, D-55099 Mainz, Germany}
\author{Hans C. Andersen\cite{hca}}
\address{Department of Chemistry, Stanford University, Stanford,
California 94305}
\maketitle

\begin{abstract}
We have performed a molecular dynamics computer simulation of a
supercooled binary Lennard-Jones system in order to compare the
dynamical behavior of this system with the predictions of the idealized
version of mode-coupling theory (MCT).  By scaling the time $t$ by the
temperature dependent $\alpha$-relaxation time $\tau(T)$, we find that
in the $\alpha$-relaxation regime $F(q,t)$ and $F_s(q,t)$, the coherent
and incoherent intermediate scattering functions, for different
temperatures each follows a $q$-dependent master curve as a function of
scaled time.  We show that during the early part of the
$\alpha$-relaxation, which is equivalent to the late part of the
$\beta$-relaxation, these master curves are well approximated by the
master curve predicted by MCT for the $\beta$-relaxation. This part is
also fitted well by a power-law, the so-called von Schweidler law. We
show that the effective exponent $b'$ of this power-law depends on the
wave vector $q$ if $q$ is varied over a large range. The early part of
the $\beta$-relaxation regime does not show the critical decay
predicted by MCT.  The $q$-dependence of the nonergodicity parameter
for $F_{s}(q,t)$ and $F(q,t)$ are in qualitative agreement with MCT.
On the time scale of the late $\alpha$-relaxation the correlation
functions show a Kohlrausch-Williams-Watt behavior (KWW).  The KWW
exponent $\beta$ is significantly different from the effective von
Schweidler exponent $b'$.  At low temperatures the $\alpha$-relaxation
time $\tau(T)$ shows a power-law behavior with a critical temperature
that is the same as the one found previously for the diffusion constant
[Phys. Rev.  Lett.  {\bf 73}, 1376 (1994)]. The critical exponent of
this power-law and the von Schweidler exponent $b'$ fulfill the
connection proposed by MCT between these two quantities. We also show
that the $q$-dependent relaxation times extracted from the correlation
functions are in accordance with the $\alpha$-scale universality
proposed by MCT.  The dynamic susceptibility $\chi''(\omega)$ data for
different temperatures also fall on a master curve when frequency is
scaled by the location of the minimum between the microscopic peak and
the $\alpha$ peak and $\chi''$ is scaled by its value at this
minimum.  The low frequency part of this master curve can be fitted
well with a functional form predicted by MCT. However, the optimal
value for the exponent parameter from this fit does not agree with the
one determined from the corresponding fit in the time domain. The high
frequency part of the master curve of $\chi''(\omega)$ cannot be fitted
well by the functional forms predicted by MCT, in accordance with our
findings from the time domain.  We test various scaling laws predicted
by the theory and find that they are qualitatively correct but that the
exponents do not fulfill certain relations predicted by the theory if
they involve the critical exponent $a$ of MCT. This discrepancy can be
rationalized by means of the strong influence of the microscopic
dynamics on the $\beta$-relaxation at early times. Those scaling laws
that do not involve the critical exponent $a$ are in qualitative and
quantitative accordance with the theory.
\end{abstract}

%
%
\section{Introduction}
\label{sec:I}
In a recent paper \cite{kob94c} we have reported some of the results we
obtained from a large scale simulation of a supercooled binary
Lennard-Jones mixture.  The aim of this work was to test whether
mode-coupling theory (MCT) is able to correctly describe the dynamical
behavior of such a system. This theory was originally developed
to describe the dynamics of simple liquids in the supercooled state
\cite{bgs84,lh84}.  However, in recent years the theory has also been
successfully applied to rationalize the dynamics of more complex
liquids. Despite these successes, there is still a great deal of
controversy on whether the theory is really able to correctly describe
the dynamics of liquids at low temperatures.  The reader can find good
introductions to the theory in some review articles
\cite{bibles,schilling94} and a list of most relevant references on MCT
in \cite{kob94c,bibles,schilling94}. Recently also a useful collection
of review articles on MCT has appeared \cite{yip95}.\par

In Ref.~\cite{kob94c} we mainly concentrated on the investigation of
the mean squared displacement of a tagged particle, the diffusion
constant, and the van Hove correlation function. We showed that at low
temperatures the mean squared displacement showed a plateau in a time
range that extended over several decades in time and could be
identified with the $\beta$-relaxation regime (see Sec.~\ref{sec:II}
for a definition of this term) predicted by MCT. At low temperatures
the diffusion constant showed a power-law behavior, in accordance with
the theory. With the help of the van Hove correlation function we
showed that the so-called cage effect is indeed present at these
temperatures. These correlation functions allowed us also to show that
the factorization property predicted by the theory holds for this
system. Furthermore we gave evidence that for this system the so-called
hopping processes are not important in the temperature range we
investigated and that therefore the dynamics of the system can be
tested with the {\em idealized} version of the theory, in which such
processes are neglected.\par

In this work we extend our analysis of the dynamical behavior of our
system to the investigation of the intermediate scattering function and
the dynamical susceptibility. Since several of the predictions of the
theory can most conveniently be tested with these quantities, this
investigation will allow us to perform more extensive tests of the
theory and thus help to decide whether MCT is able to correctly
describe the dynamical behavior of simple supercooled liquids. Some of
these results have been reported already in a previous paper
\cite{kob94a} where we presented the scaling behavior of the incoherent
intermediate scattering function for a value of the wave vector $q$ in
the vicinity of the maximum of the structure factor.  These results are
here extended to a larger range of values of $q$ and different types of
correlation functions.\par

The rest of the paper is organized in the following way: In Sec.~II
we give a short review of some of the predictions of the theory in
order to facilitate the understanding of the subsequent tests of the
theory presented in this work. In Sec.~III we introduce some of the
details of our model and of our simulation. Section~IV is then
devoted to the presentation of our results, which are summarized in
Sec.~V.

\section{Mode Coupling Theory}
\label{sec:II}
In order to facilitate the reading of this paper, we compile in this
section some of the predictions of MCT. The derivation of these
predictions can be found in the original papers or in review
articles \cite{bibles,schilling94}.\par

Mode-coupling theory attempts to describe the dynamics of strongly
supercooled liquids at temperatures slightly above the glass transition
temperature $T_{g}$.  In its simplest version, the so-called idealized
MCT, the theory predicts the existence of a temperature $T_{c}$ at
which the system undergoes a transition from ergodic behavior to
nonergodic behavior. This means that certain types of correlation
functions, as, e.g., the intermediate scattering function for wave vector
$q$, $F(q,t)$, do not decay to zero even for long times if $T<T_{c}$.
This transition is predicted to be observable for all time correlation
functions $\langle X(0)Y(t) \rangle$ between dynamical variables $X$
and $Y$ for which the overlap with the Fourier transform of the density
fluctuations $\delta \rho(q)=\rho(q)-\langle \rho(q) \rangle$ is
nonzero, i.e.  for which $\langle \delta \rho (q) X\rangle\neq 0$
and $\langle \delta \rho (q) Y\rangle \neq 0$. Here $\langle
\rangle$ stands for the canonical average. The following results are
all of an asymptotic nature in the sense that they are valid only if
$\epsilon\equiv (T-T_{c})/T_{c}$, the small parameter of the theory,
tends to zero.\par

The theory also predicts the existence of a parameter $\lambda$, the
so-called exponent parameter, which is very important for the {\em
quantitative} description of the relaxation behavior (see below for
details).  This exponent parameter can be computed if the structure
factor of the system is known with sufficient precision, but since this
is seldom the case it is usually treated as an adjustable parameter for
fitting the data.\par

Consider a normalized time correlation function $\phi(t)=\langle X(0)
Y(t) \rangle / \langle XY \rangle $ for two dynamical variables $X$ and
$Y$ that have a nonvanishing overlap with $\delta \rho (q)$.  MCT
predicts that for temperatures just above $T_{c}$, $\phi(t)$ shows a
two step relaxation behavior. Starting with the value of unity at time
zero $\phi(t)$ is supposed to decay to a value of $f_{c}>0$, the
so-called nonergodicity parameter, and then decay slowly to zero for
long times.  Thus if $\phi(t)$ is plotted versus the {\em logarithm} of
time, the function quickly decays from unity on a microscopic time
scale, then slowly decays to a plateau of height $f_{c}$, and finally
slowly decays to zero.  In the language of MCT the time range for which
$\phi(t)$ is close to this plateau is called the $\beta$-regime and the
time range that starts with the correlators beginning to deviate from
this plateau and that extends to infinite time is called the
$\alpha$-regime. Two things should be noted in order to avoid of
getting confused. The first is that the $\beta$-relaxation regime of MCT
should not be confused with the $\beta$-relaxation as it has been
introduced by Johari and Goldstein \cite{joharietal} since the latter
is a {\em peak} in the spectrum whereas the former corresponds to a
{\em minimum} in the spectrum. The second thing to note is that the
{\em late} part of the $\beta$-regime overlaps with the {\em early}
part of the $\alpha$-regime. Thus the two regimes should not be
considered as {\em completely} unrelated relaxation regimes.\par

For times in the $\beta$-relaxation regime MCT predicts that the
correlator $\phi(t)$ can be written in the following form:
\begin{equation}
\phi(t)=f_{c}+hG(t).
\label{eq1}
\end{equation}
Here $f_{c}$ is the above mentioned nonergodicity parameter and $h$ is
a positive amplitude factor. Both quantities will depend on the nature of
$\phi$, e.g. they will depend on the wave vector $q$ if $\phi$ is the
intermediate scattering function, but not on temperature $T$ or time
$t$. The whole temperature and time dependence of the right hand side
of Eq.~(\ref{eq1}) is given by the function $G(t)$. This
function is predicted by MCT to be of the form:
\begin{equation}
G(t)=\sqrt{|\epsilon|}g(t/t_{\epsilon}).
\label{eq2}
\end{equation}
where $\epsilon$ is the small parameter of the theory introduced above
and $t_{\epsilon}$ is the time scale of the $\beta$-relaxation.  This
time scale is predicted to show a power-law dependence on $T$ with a
divergence at $T=T_{c}$:
\begin{equation}
t_{\epsilon}\sim (T-T_{c})^{-1/2a} \quad ,
\label{eq2n}
\end{equation}
where the quantity $a$ can be computed from the exponent parameter
$\lambda$ (see below). If $\lambda$ is known, the function
$g(t/t_{\epsilon})$ in Eq.~(\ref{eq2}) can be computed explicitly. It
has been shown analytically that for times much larger than the
microscopic times $t_{0}$ but less than $t_{\epsilon}$, $g(t)$ is a
power-law, in this context often called the critical decay, i.e. it is
of the form:
\begin{equation}
g(t/t_{\epsilon})=(t_{\epsilon}/t)^{a} \quad,\quad t_{0}\ll t \leq
t_{\epsilon}.
\label{eq3}
\end{equation}
Here the exponent $a$, often also called the critical exponent, is the
same quantity that appeared in Eq.~(\ref{eq2n}).  \par

For times $t$ that are larger than $t_{\epsilon}$ but much smaller than
the $\alpha$-relaxation time $\tau$, $g(t/t_{\epsilon})$ is predicted to
be also a power-law, which in this context is usually called the von
Schweidler law, i.e.
\begin{equation}
g(t/t_{\epsilon})=-B(t/t_{\epsilon})^{b} \quad,\quad t_{\epsilon}\leq t
\ll \tau .
\label{eq4}
\end{equation}
Here the prefactor $B$ and the exponent $b$, the so-called von
Schweidler exponent, can also be computed when the exponent parameter
$\lambda$ is know. In particular MCT predicts that the exponent $a$ of
the critical decay and the von Schweidler exponent $b$ are related to
the exponent parameter via:
\begin{equation}
\lambda=\frac{\Gamma(1-a)^{2}}{\Gamma(1-2a)}=
\frac{\Gamma(1+b)^{2}}{\Gamma(1+2b)}\quad,
\label{eq5}
\end{equation}
where $\Gamma(x)$ is the $\Gamma$-function.\par

For times on the time scale of the $\alpha$-relaxation regime MCT
predicts that the correlator $\phi(t)$ obeys the so-called time
temperature superposition principle. This means that correlators for
different temperatures (all close to $T_c$, of course) will fall onto a
master curve if time is scaled by the $\alpha$-relaxation time $\tau$,
i.e.
\begin{equation}
\phi(t)=F(t/\tau(T)).
\label{eq5n}
\end{equation}
For the late $\alpha$-relaxation regime the master function $F(t/\tau)$
is predicted to be well approximated by a Kohlrausch-Williams-Watt
function (KWW), often also called stretched exponential, i.e.
\begin{equation}
\phi(t)\approx A\exp\left(-(t/\tau)^{\beta}\right),
\label{eq6}
\end{equation}
and recently it has been shown that if $\phi(t)$ is the intermediate
scattering function for wave-vector $q$ of a simple liquid
Eq.~(\ref{eq6}) becomes exact for large values of $q$
\cite{fuchs94,fuchsal93}.  The $\alpha$-relaxation time $\tau$ will in
general depend on the specific nature of $\phi$, e.g. on the wave
vector $q$ if $\phi$ is the intermediate scattering function $F(q,t)$.
However, MCT predicts that the relaxation times of all correlators
should show a divergent behavior near $T_{c}$ in the form of a
power-law with an exponent $\gamma$ that is independent of the type of
correlator studied. Thus the relaxation time of the intermediate
scattering function $F(q,t)$ is predicted to be of the form
\begin{equation}
\tau(q)=C(q)(T-T_{c})^{-\gamma}.
\label{eq7}
\end{equation}
Here $C(q)$ is a smooth function of temperature. Thus in the vicinity
of $T_{c}$ the main dependence of $\tau$ on temperature is given by
the power-law behavior in Eq.~(\ref{eq7}). This property is called the
$\alpha$-scale universality. The exponent $\gamma$ in Eq.~(\ref{eq7})
can be computed once $\lambda$ is known by means of
\begin{equation}
\gamma=\frac{1}{2a}+\frac{1}{2b}\quad,
\label{eq8}
\end{equation}
where $a$ and $b$ are the two exponents from Eq.~(\ref{eq5}).\par

By making use of equations (\ref{eq1})-(\ref{eq2n}), (\ref{eq4}) and
(\ref{eq7}) it is simple to show that on the time scale of the late
$\beta$-relaxation regime, i.e. the time scale for which the von
Schweidler law in Eq.~(\ref{eq4}) is predicted to hold, the von
Schweidler law can also be written as follows:
\begin{equation}
\phi(t)=f_{c}-hB(t/\tau)^{b}\quad,
\label{eq9}
\end{equation}
where $h$ and $B$ are the temperature independent constants of
Eqs.~(\ref{eq1}) and (\ref{eq4}). Note that MCT predicts that the
exponent $b$ in Eqs. (\ref{eq4}), (\ref{eq5}), (\ref{eq8}) and
(\ref{eq9}) is in general not the same as the KWW exponent $\beta$ in
Eq.~(\ref{eq6}).  Thus Eq.~(\ref{eq9}) is {\it not} the short time
expansion of the KWW law in Eq.~(\ref{eq6}).\par

If the correlation function $\phi(t)$ is time Fourier transformed and
multiplied by the frequency $\omega$ one obtains the dynamic
susceptibility $\chi''(\omega)$. Since $\phi(t)$ is predicted to show,
at temperatures just above $T_{c}$, a two step relaxation behavior,
$\chi''(\omega)$ is predicted to show a double peak structure at these
temperatures. MCT makes predictions about the following quantities:
$\omega_{\epsilon}$, the frequency at the minimum between the two peaks;
$\chi^{\prime\prime}_{\epsilon}\equiv\chi^{\prime\prime}(\omega_{\epsilon})$,
the value of the susceptibility at this minimum; and $\omega_{max}$, the
frequency of the peak that occurs at lower frequency (the so-called
$\alpha$-peak).  In particular MCT predicts the two following power-law
dependencies for $\chi_{\epsilon}''$ and $\omega_{max}$ on temperature:
\begin{equation}
\chi_{\epsilon}'' \sim (T-T_{c})^{1/2}
\label{eq10}
\end{equation}
and
\begin{equation}
\omega_{max}\sim \tau^{-1} \sim (T-T_{c})^{\gamma}\quad,
\label{eq11}
\end{equation}
where we made use of Eq.~(\ref{eq7}). Making use of Eqs.~(\ref{eq2n})
and (\ref{eq10}) it follows that
\begin{equation}
\chi_{\epsilon}''\sim \omega_{\epsilon}^{a}
\label{eq12}
\end{equation}
and using Eqs.~(\ref{eq2n}), (\ref{eq8}) and (\ref{eq11}) one can show
that
\begin{equation}
\omega_{\epsilon} \sim \omega_{max}^{b/(a+b)}.
\label{eq13}
\end{equation}

It should be recognized that Eq.~(\ref{eq13}) can be derived from
Eqs.~(\ref{eq2n}), (\ref{eq8}) and (\ref{eq11}) even if Eq.~(\ref{eq5})
does not hold.  This point will be important later when we discuss our
results. \par

Note that some of these predictions are consequences of the simplest
version of MCT, the so-called idealized MCT. If thermally activated
processes are present, in this context usually called hopping
processes, some of the above predictions have to be modified. However,
in a previous paper we have given evidence that for our system these
hopping processes are not important in the temperature range we
investigated \cite{kob94c}. Thus it is appropriate to compare the
low-temperature dynamics of our system with the idealized version of
the theory.\par

\section{Model and Simulation}
\label{sec:III}
In this paper we give only some of the most important details of the
model investigated and the essential features of the simulation. More
details can be found in Ref.~\cite{kob94c}.\par

The system considered is a binary mixture of Lennard-Jones particles.
Both types of particles (type A and type B) have the same mass $m$. The
interaction potential $V_{\alpha\beta}(r)$ is given by
$V_{\alpha\beta}(r)=4\epsilon_{\alpha\beta}
[(\sigma_{\alpha\beta}/r)^{12}-(\sigma_{\alpha\beta} /r)^{6}]$
($\alpha,\beta \in \{A,B\}$) with the following set of parameters:
$\epsilon_{AA}=1.0$, $\epsilon_{AB}=1.5$, $\epsilon_{BB}=0.5$,
$\sigma_{AA}=1.0$, $\sigma_{AB}=0.8$, and $\sigma_{BB}=0.88$. For
computational efficiency these potentials were truncated and shifted at
a distance of 2.5$\sigma_{\alpha\beta}$. In the following we report all
quantities in reduced units, i.e. length in units of $\sigma_{AA}$,
energy in units of $\epsilon_{AA}$ and time in units of
$(m\sigma_{AA}^2/48\epsilon_{AA})^{1/2}$. For Argon these units
correspond to a length of 3.4\AA, an energy of 120$Kk_B$ and a time of
$3\cdot10^{-13}$s.\par

The number of A and B particles was 800 and 200, respectively. The
equations of motions were solved with the velocity form of the Verlet
algorithm with a time step of 0.01 and 0.02 at high ($T\geq 1.0$) and
low ($T<1.0$) temperatures, respectively. The length of the cubic box
was $L=9.4\sigma_{AA}$ and periodic boundary conditions were applied.
The system was constructed at a high temperature ($T=5.0$) and
subsequently cooled to lower temperatures by coupling it to a
stochastic heat bath.  The temperatures investigated were $T=5.0$, 4.0,
3.0, 2.0, 1.0, 0.8, 0.6, 0.55, 0.50, 0.475, and 0.466. Multiple runs
were performed at each temperature, and the correlation functions were
averaged. Since MCT assumes the system to be in equilibrium, great care
was taken in order to make sure that we gave the system enough time to
equilibrate at {\em all} temperatures. This was done by allowing the
system to equilibrate at each temperature for a time that was longer
than the time of the $\alpha$-relaxation.  In Ref.~\cite{kob94c} we
give strong evidence that the cooling schedule and the equilibration
times are such that the dynamical behavior observed is really that of
an equilibrium system.\par

\section{Results}
\label{sec:IV}
In this section we present the results of our simulation. In the first
part we will report our findings about the intermediate scattering
function. This quantity is very useful in order to test certain kind of
predictions of MCT. However, other predictions of the theory are more
easily tested with the help of the dynamic susceptibility and
therefore we devote the second part of the section to the investigation
of this quantity.\par

\subsection{Intermediate Scattering Function}
\label{ssec:IV-A}

The intermediate scattering function is given by the space Fourier
transform of the van Hove correlation function $G(r,t)$
\cite{hansenmcdonald86}. The latter one can be split into two parts,
the self part $G_{s}(r,t)$, and the distinct part $G_{d}(r,t)$.
Consequently there exists also two different types of intermediate
scattering functions: $F_{s}(q,t)$, the incoherent intermediate
scattering function, which is the Fourier transform of $G_{s}(r,t)$, and
$F(q,t)$, the coherent intermediate scattering function, which is the
Fourier transform of the van Hove correlation function $G(r,t)$.  We
computed these two correlation functions by computing first the thermal
average of the van Hove correlation function and then took the Fourier
transform of this function\cite{footnote1}. Although this procedure is
formally exact in the thermodynamic limit it turned out not to be the
best way to compute the intermediate scattering function. The problem
is that at low temperatures the distinct part of the van Hove
correlation function does not attain its asymptotic value, i.e.
unity, at the distance $L/2$, i.e.  at the largest distance accessible
in the simulation ($L$ is the length of the box). This can be seen,
e.g., in Fig.~10 of Ref.~\cite{kob94c}.  Therefore the Fourier
transform of this quantity shows some oscillations with a periodicity
(in $q$) of $4\pi/L$. A possible solution of this problem would have
been to compute the intermediate scattering function from the positions
of the particles and only afterwards do the average over different
configurations; unfortunately, this problem was recognized only after
the simulations had been performed.  However, we believe that the
oscillations are only a minor flaw in the data and in particular do not
obscure the relaxation behavior of the correlation functions.\par

In Fig.~\ref{fig1} we show the structure factors $S(q)$, given by
$F(q,t)$ at $t=0$, versus $q$ for all temperatures investigated. For
clarity the curves for the different temperatures have been displaced
vertically (see figure caption for details). At low temperatures (top
curves) we clearly see the oscillations mentioned in the previous paragraph.
They are most pronounced for small values of $q$ and hardly noticeable
for large values of $q$. In the following we will mainly concentrate to
the range of $q$ that is between the first maximum and the first
minimum in $S(q)$. As can be recognized from the figure in this range
the amplitude of the oscillations is small and thus this effect can
probably be neglected.\par

We recognize from Fig.~\ref{fig1} that $q_{max}(T)$, the location of the
first maximum in $S(q)$, depends only very weakly on temperature. The
same holds for $q_{min}(T)$, the first minimum in $S(q)$. In the following
we will, among other things, study the temperature dependence of the
correlation functions at $q=q_{max}$ and $q=q_{min}$. Since the
temperature dependence of $q_{max}$ and $q_{min}$ is so weak we will
neglect it altogether and fix the values of the two quantities to their
corresponding values at the lowest temperature. From the same figure we
can also see that at low temperatures the form of $S(q)$ depends only
weakly on temperature.  The only thing that changes at these
temperatures is that the height of the peaks and the depth of the
valleys become {\em slightly} more pronounced. In Ref.~\cite{kob94c}
and \cite{kob94a} we have found that at low temperatures the relaxation
times show a strong temperature dependence.  Since we recognize now
that this slowing down is not accompanied by an significant change in
the structure factor we can exclude the possibility that the slowing
down is associated with some sort of divergence of the correlation
length of the pair-correlation function. This observation is in
accordance with the underlying idea of MCT that the slowing down of the
dynamics is a purely dynamic phenomenon and has nothing to do with
some sort of underlying phase transition. However, one has of course to
keep in mind that the latter possibility is by no means excluded by our
observation of the independence of $S(q)$ on temperature, since it may
well be that a {\em different} quantity indeed shows a diverging
correlation length.\par

In Fig.~\ref{fig2} we show the incoherent intermediate scattering function
$F_{s}(q,t)$ for all temperatures investigated. Figures (a) and (b) are
for the A particles with $q=q_{max}=7.25$ and $q=q_{min}=9.61$,
respectively, and figures (c) and (d) are for the B particles for
$q=q_{max}=5.75$ and $q=q_{min}=7.06$, respectively.  We see that for
{\em all} temperatures investigated the correlation functions decay to
zero in the long time limit. This means that the fluctuations that
were present at time zero have disappeared in the time span of the
simulation. Thus this is evidence that the length of the simulation is
large enough that the system can come to equilibrium at all
temperatures. Other evidence for this was presented in
Ref.~\cite{kob94c}.\par

For short times the correlators show a quadratic dependence on time,
which can be understood by remembering that for short times the motion
of the particles is essentially ballistic. For intermediate and long
times the correlators at high temperatures (curves to the left) show a
relaxation behavior that is similar to a simple exponential decay.
This behavior changes when we go to intermediate temperatures
($T\approx 1.0$). There we see that for intermediate times a small
shoulder begins to form. This temperature is comparable to the one for
which the diffusion constant $D$ and the relaxation times $\tau$
started to show \cite{kob94c,kob94a} the asymptotic behavior at low
temperature predicted by MCT, i.e. a power-law with critical
temperature $T_{c}$ and critical exponent $\gamma$ (see
Eq.~(\ref{eq7})). Thus the qualitative change in the relaxation
behavior of the intermediate scattering function, e.g. the occurrence
of a shoulder, is accompanied with the onset of the asymptotic behavior
in $D$ and $\tau$.\par

When the temperature is lowered even further this small shoulder
becomes more pronounced until we observe almost a plateau at the lowest
temperature.  Thus we find that at low temperatures the correlators
exhibit the two-step relaxation phenomenon predicted by MCT. We also
note that at low temperatures the correlators for the A particles (Fig.
\ref{fig2}a and \ref{fig2}b) show a small bump for times around 14 time
units. A similar phenomena was observed in a recent computer simulation
of Lewis and Wahnstr\"om of orthoterphenyl\cite{lewisgw}. In that work
evidence was given that this bump is a finite size effect. A similar
bump was also observed in a simulation of a different Lennard-Jones
mixture\cite{wahnstrom91}, a simulation of a molten
salt\cite{signorinijlbmlk90} and a simulation of a colloidal suspension
\cite{lowen91}. However, no such feature was observed in simulations
with soft spheres\cite{softsphnobump}. \par

A comparison of the correlators plotted in Fig.~\ref{fig2}a with those
in Fig.~\ref{fig2}b (or of Fig.~\ref{fig2}c with Fig.~\ref{fig2}d) shows
that the height of the plateau as well as the time scale on which the
correlator ultimately decays to zero depend on the value of $q$. In
order to show these two effects clearer we show in Fig.~\ref{fig3}
$F_s(q,t)$ for the A particles at $T=0.466$, the lowest temperature
investigated, for values of $q$ between $6.0\leq q \leq 14.0$. From this
figure it becomes evident that the height of the plateau depends
strongly on the value of $q$ and that also the relaxation time varies
by about one order of magnitude in this range of $q$. MCT predicts the
qualitative dependence of this height on $q$ and later on we will
present the result of our analyses of this dependence in more detail.
Also the investigation of the dependence of the relaxation time on $q$
will be postponed for the moment.\par

MCT predicts that for low temperatures the correlators should obey the
time temperature superposition principle in the $\alpha$-relaxation
regime, i.e.  show a scaling behavior if they are plotted versus
rescaled time $t/\tau$, where $\tau(T)$ is the relaxation time for the
$\alpha$-relaxation, (see Eq.~(\ref{eq5n})). To test this prediction of
the theory we made such a scaling plot.  We defined the
$\alpha$-relaxation time of a correlator as the time where the
correlator has decayed to $e^{-1}$ of its initial value. Note that this
kind of arbitrary definition of the relaxation time makes, within the
framework of MCT, perfect sense since, due to the time temperature
superposition principle (Eq.~(\ref{eq5n})), {\em any} definition of a
relaxation time that measures the time scale of the
$\alpha$-relaxation, is predicted to show the same temperature
dependence.  In Fig.~\ref{fig4} we show the incoherent intermediate
scattering function $F_s(q,t)$ versus this rescaled time for the A and
B particles and for $q=q_{min}$ and $q=q_{max}$. From these figures we
recognize that at low temperatures we find indeed a scaling behavior
thus confirming this prediction of the theory.\par

MCT predicts the functional form of the master function $G(t)$ in the
$\beta$-relaxation regime (see Eq.~(\ref{eq2})). Thus we tried to fit
our master curve with the one predicted by the theory and the best fit
we obtained is included in the figures as well (dashed line)
\cite{footnote2}. The value of the exponent parameter $\lambda$ that
gave the best fit is given in each figure. We recognize that this fit
is very good for rescaled times in the interval $10^{-3} \leq t/\tau
\leq 10^{0}$, thus over a time range spanning about three decades.
This time range corresponds to the {\em late} $\beta$-regime which is
the same as the {\em early} $\alpha$-regime. From this we conclude that
MCT is able to rationalize the master curve in the mentioned relaxation
regime.\par

The theory predicts that the exponent parameter is independent of the
type of correlator or the value of $q$. The values of $\lambda$ that we
obtained are not all equal, but instead have a variation of about 5\%
for the four correlators. Since the statistical uncertainty with which
the fitting procedure can determine any value of $\lambda$ is about 1\%
we thus find that for our system the exponent parameter is not
constant. However, it should be remembered that the prediction of MCT,
that $\lambda$ is independent of the type of correlator or the value of
$q$, is an {\em asymptotic} result of the solutions of equations that
are only an approximation to the original MC equations, i.e. the
equations in which the full $q$ dependence is taken into account. Thus
it can be expected that there will be corrections to these asymptotic
results. Furthermore, the values of $\lambda$ we determined depend to
some extent on the time range where the fit to determine $\lambda$ was
done and thus an additional, systematic, error might be introduced
which we estimate to be of the order of several percent.  Thus a small
dependence of $\lambda$ on $q$ or the type of correlator should not be
viewed as a failure of the theory.\par

In order to test this prediction of MCT more extensively we present in
Fig.~\ref{fig5} the coherent part of the intermediate scattering
function for the AA, the AB and the BB correlation. The values of $q$
were chosen to be at the maximum of the corresponding structure
factors. From this figures we recognize that, in accordance with MCT,
also these correlation functions show a scaling behavior in the
$\alpha$-relaxation regime. In the late $\beta$-relaxation regime the
master curves can again be fitted very well by functional forms
predicted by MCT (dashed lines).  Thus we find that this prediction of
the theory holds also for these types of correlation function. The
values of the exponent parameter $\lambda$ for the different
correlators are similar to the ones we presented in Fig.~\ref{fig4}.
Thus we can conclude from these two sets of figures that $\lambda$ is
indeed only a weak function of $q$ for those values of $q$ studied or
the type of correlator investigated.\par

For times on the time scale of the late $\beta$-relaxation regime MCT
predicts that the theoretical master function is given by a power-law,
the so-called von Schweidler law (see Eq.~(\ref{eq4})). In previous
work we have shown that the master curve of our data can be fitted very
well with such a functional form \cite{kob94a,kob94b}. In these papers
the von Schweidler exponent $b$ was treated as a fit parameter and it
was found that $b$ did neither depend strongly on the type of correlator
nor the value of $q$. Because of the one-to-one connection between the
von Schweidler exponent $b$ and the exponent parameter $\lambda$, see
Eq.~(\ref{eq5}), this can be seen as evidence that the exponent
parameter is almost constant for our system. This observation is
therefore in accordance with the findings presented here. The fits with
the von Schweidler law are included in figures~\ref{fig4}
and~\ref{fig5} as well, and we recognize that in the region where the
theoretical master curve of MCT is fitting the data well the von
Schweidler law does so too. Thus we can conclude that for our system
the power-law is a good approximation to the theoretical master curve
for the whole time range of the late $\beta$-relaxation and not only
very close to the plateau as might be expected a priori from the
asymptotic nature of the von Schweidler law.  Since we have determined
the exponent parameter for each of the correlation functions shown, we
can compute the von Schweidler exponent $b$ by means of Eq.~(\ref{eq5})
and compare it with the result from fitting only the power-law. In
order to distinguish these two quantities we will denote the latter by
$b'$. The values of these quantities are given in figures \ref{fig4}
and \ref{fig5}.  We find that $b$ and $b'$ are usually very close to
each other and can be taken essentially to be equal. Therefore we will
report in the following only the value of the von Schweidler exponent
as determined from the power-law fit.\par

For times belonging to the late part of the $\alpha$-relaxation regime,
one can show that the master function predicted by MCT is well
approximated by a Kohlrausch-Williams-Watt law (see Eq.~(\ref{eq6})).
We have tried to fit this part of the master curve of our data with
such a functional form and the result is included in figures~\ref{fig4}
and~\ref{fig5} as well. Although it is not clearly seen in figures on
this scale, these fits are very good at low temperatures and long times
and therefore we can conclude that this prediction of the theory holds
too. Note that the values of the KWW exponents $\beta$ are
significantly different from the ones of the von Schweidler exponent
$b'$. Thus it is {\em not} the case that the von Schweidler law can be
considered as the short time expansion of the KWW law. One might be
tempted to try to fit the {\em whole} master curve with a KWW law, i.e.
the whole $\alpha$-relaxation regime. However, we found that such a fit
is not convincing at all and can therefore be ruled out. \par

Besides the von Schweidler law, which MCT predicts to be present for
the late $\beta$-relaxation regime, the theory also predicts that the
correlation functions should show a power-law, the so-called critical
decay, also when {\em approaching} the plateau (see Eq.~(\ref{eq3})).
We have thus tried to fit the data in the early $\beta$-relaxation
regime with such a functional form but were not able to find any sign
of the presence of such a relaxation behavior in this time regime.
Inspection of figures~\ref{fig4} and \ref{fig5} shows that this apparent
absence of the critical decay is probably due to the fact that on the
time scale at which the critical decay is supposed to be present the
correlators seem to be still strongly influenced by the relaxation
behavior at short times, where the dynamics is essentially ballistic.
This influence will thus make the detection of a critical decay very
difficult. We will come back to this pint in the next section when we
discuss the dynamic susceptibility.\par

It is interesting to compare our results with the ones that Bengtzelius
obtained from a numerical integration of the mode-coupling equations
for a monoatomic Lennard-Jones system \cite{bengtzelius86}. These
results were later improved by Smolej and Hahn \cite{smolej93}. By
taking into account essentially the full $q$ dependence of the
structure factor and also making a reasonable modelling of the short
time dynamics of this system, these authors obtained the full time
dependence of the intermediate scattering function. Their results show
that the correlation functions are qualitatively very similar to the
ones presented in this work. In particular they also show that in the
time domain the critical decay is hardly noticeable if the system is
not very close to the critical point.  In Ref.~\cite{kob94c} and
\cite{kob94a} we have shown that at the lowest temperature we
investigate in this simulation $\epsilon=(T-T_{c})/T_{c}$ is about
0.07. In order to compare this value of $\epsilon$ to the ones studied
in Ref.~\cite{bengtzelius86} one has to remember that Bengtzelius
varied the density of the system and not the temperature, as we have
done it in this work. Thus in the work of Bengtzelius we have
$\epsilon=(n_c-n)/n_c$, where $n$ is the particle density.  Thus the
meaning of the small parameter $\epsilon$ in his and our work is
clearly not the same. However, we can compare the relaxation behavior
of the incoherent part of the intermediate scattering function $F(q,t)$
for $q$ close to the first peak in the structure factor, presented in
Fig.~3 of Ref.~\cite{bengtzelius86}, with the corresponding result of
our work.  (In Fig.~\ref{fig2}a we show the relaxation behavior of
$F_s(q,t)$ which is qualitatively very similar to the one of $F(q,t)$.)
{}From this comparison we conclude that the relaxation behavior at our
lowest temperature ($\epsilon=0.07$) corresponds to a value of
$\epsilon$ of about 0.0023 in the work of Bengtzelius (curve C of
Fig.~3 in Ref.~\cite{bengtzelius86}). Smolej and Hahn have shown that in
the monoatomic Lennard-Jones system the critical decay can be observed
only for $|\epsilon| \approx 0.00042$, thus about a factor of 5
(=0.0023/0.00042) smaller than the ones accessible in our simulation.
Thus if we also take into account that our correlators have a certain
amount of noise it is very reasonable that the critical decay becomes
completely obscured.  Thus we conclude that, for this system and the
temperature range investigated the critical decay is either not present
at all or not detectable with data of the accuracy we are able to
obtain.\par

We turn now our attention to the nonergodicity parameter $f_{c}$. As we
already noted in Fig.~\ref{fig3}, $f_{c}$ depends on $q$. MCT predicts
that for the incoherent intermediate scattering function the
nonergodicity parameter shows a Gaussian-like behavior as a function of
$q$. For the coherent intermediate scattering function $f_{c}$ is
predicted to show an oscillatory behavior with oscillations that are in
phase with the structure factor $S(q)$. In order to test these
predictions we fit the correlators $F_{s}(q,t)$ and $F(q,t)$ at the
lowest temperature investigated with a von Schweidler law and used the
offset as an approximation to the nonergodicity parameter. The quality
of this approximation was tested for the cases presented in
figures~\ref{fig4} and \ref{fig5} in that we compared the offset we
obtained from the power-law fit with the nonergodicity parameter we
obtained when we fitted the full theoretical master curve. In most
cases investigated the difference between the two quantities was less
than 1\% and thus this approximation should be of sufficient
accuracy.\par

In Fig.~\ref{fig6} we show the $q$-dependence of the nonergodicity
parameters $f_{c}$ for the incoherent and coherent intermediate
scattering function for the A particles. From this figure we recognize
that the $f_{c}$ for the incoherent part shows indeed a Gaussian-like
behavior (upper dotted curve). The nonergodicity parameter
of the coherent part (upper solid curve) shows an
oscillatory behavior that is in phase with the structure factor $S(q)$
(dashed curve).  Also the relative magnitude of the two curves is very
similar to the one predicted by MCT \cite{bgs84}. Thus we conclude
that the prediction of MCT concerning the $q$-dependence of the
nonergodicity parameter of the intermediate scattering function is in
qualitative accordance with our data. A similar accordance with this
prediction of the theory was reported also from scattering
experiments \cite{megen94} and another computer simulation
\cite{barrat90}.\par

Also included in the figure are the amplitudes $A$ of the KWW law we
fitted at long times for the incoherent intermediate scattering
function (lower dotted curve) and the coherent one (lower
solid curve). We recognize that these amplitudes show
qualitatively the same behavior as the nonergodicity parameters and can
thus serve as a {\em first} approximation to it.  However, we also note
that both KWW amplitudes are always smaller that the corresponding
nonergodicity parameters.  Thus this is evidence that the
KWW law which fits the data well on the time scale of the late
$\alpha$-relaxation does not fit the data well on the time scale of
the early $\alpha$-relaxation.\par

MCT predicts that the exponent parameter depends only on the structure
factor at the transition. From this it follows that the von Schweidler
exponent $b$ is independent of the correlator studied, since the two
quantities are related by means of Eq.~(\ref{eq5}). In figures
\ref{fig4} and \ref{fig5} we gave evidence that the exponent parameter
$\lambda$ does not depend strongly on the type of correlator or the
value of $q$. In order to make a more systematic test of this
prediction of MCT we determined the von Schweidler exponent $b'$ for
various correlation functions and varied $q$ over a large range.
Similar to the nonergodicity parameter, $b'$ was determined by means of
a power-law fit to the correlators at the lowest temperature. The range
in time for which this fit was good was about three orders of magnitude
for small values of $q$ and about two orders of magnitudes for large
values of $q$. Thus we were able to determine $b'$ with an absolute
accuracy of about 0.02. Note that the von Schweidler exponent
determined in this way is an {\em effective} von Schweidler exponent,
since the fit was done by trying to fit the power-law to the data over
a time interval which was as large as possible. Thus it might be, that
if one would restrict the time range over which the fit was done to a
smaller interval, the value of the exponent would change.  However,
since the fits were usually quite good over several orders of magnitude
in time, a restriction of the time interval would probably not lead to
very different values for the exponent.\par

In Fig.~\ref{fig7} we plot the effective von Schweidler exponent $b'$,
determined in the way explained above, versus $q$ for the incoherent
intermediate scattering function for the A and B particles as well as
for the coherent one for the AA, AB and BB correlation. From this
figure we recognize that the effective von Schweidler exponent for the
incoherent function decreases with increasing $q$. For the coherent
function, AA and BB correlation, $b'$ shows also a general trend to
decrease but we see in addition some oscillations which are in phase
with the structure factor (see Fig.~\ref{fig1}). For the AB correlation
$b'$ also shows the trend to decrease but the peaks in the curve cannot
be assigned to maxima in the corresponding structure factor. We see
that for $q$ in the range shown in the figure the variation of $b'$
between the different curves is quite appreciable and also the
systematic dependence on $q$ is clearly visible. At first glance this
seems to be in contradiction with our previous findings in which we
reported only a weak dependency of $b'$ on $q$ (see, e.g.,  Fig.~4 in
Ref.~\cite{kob94a} or the results presented in figures~\ref{fig4} and
\ref{fig5}). However, it should be noted that in Ref.~\cite{kob94a} we
focussed on the dependency of $b'$ for the incoherent part of the A
particles only in the range of $q$ between $6.5 \leq q \leq 9.6$, i.e.
the range from about $q_{max}$ to $q_{min}$. We see from
Fig.~\ref{fig7} that in this relatively small range, the value of $b'$
does indeed not change a lot and can thus be considered to be almost
constant. Only if the von Schweidler exponent is measured over a much
larger interval of $q$ is its dependence on $q$ revealed. If the values
of $b'$ for $F_{s}(q,t)$ of the A and B particles are read off at
$q=q_{max}$, where $q_{max}$ is the location of the maximum in the {\em
corresponding} structure factor, we find that these values are quite
close together.  This is in accordance with the finding presented in
the context of figures~\ref{fig4} and \ref{fig5}. Thus we draw the
conclusion from this figure that this {\em effective} von Schweidler
exponent $b'$ depends on $q$, but that its value at the maximum of the
corresponding structure factor is almost independent of the type of
correlator.\par

In Fig.~4 of Ref.~\cite{kob94a} we presented a graph in which we
plotted $F_{s}(q,t)$ for the A particles versus $t^{b'}$, where $b'$
was the effective von Schweidler exponent determined for $q=q_{max}$,
for values of $q$ ranging from a bit less than $q_{max}$ to values up
to $q_{min}$. If $F_{s}(q,t)$ is a power-law with exponent $b'$ the
curves will be straight lines. With this plot we gave evidence that the
von Schweidler exponent was essentially independent of $q$ in this
range of $q$. Since we find now that, if $q$ is varied over a larger
range, the effective von Schweidler exponent is dependent on $q$ we
tried to test if it is possible to describe at least {\em part} of the
late $\beta$-relaxation regime with a von Schweidler law with an
exponent that is independent of $q$.  MCT predicts that the time range
in which the von Schweidler law, with a $q$-independent exponent, holds
should depend on $q$.  Thus it might be that the fitting procedure
described above might make use of data over too large a range in time
and thus extract an effective exponent $b'$ that is significantly
different from the real von Schweidler exponent $b$. Thus we tried to
plot $F_{s}(q,t)$ for the A particles for $q$ in the range $2.0 \leq q
\leq 24.0$ versus $t^{0.49}$. Here the exponent $b'=0.49$ stems from
our fit for $q=q_{max}$ (see Fig.~\ref{fig4}a). Unfortunately the
resulting plot was not very useful to decide whether a constant value
of $b'$ is compatible with the data or not, since noise in the data
prevented us from determining reliably those parts of the individual
curves where they are straight lines. Only in the range between
$q_{max}$ and $q_{min}$ was the statistics good enough to identify
clearly the time range where a straight line was present and this
result was already presented in Ref.~\cite{kob94a}.  Thus the only
conclusion we can draw at the moment is that the exponent of the
power-law, if fitted over a time range as large as possible, is indeed
dependent on $q$. If there does exist a $q$-dependent time range in
which a power-law with a constant value of $b'$ fits the data well,
this range is not observable within the accuracy of our data, except
for $q$ between $q_{max}$ and $q_{min}$.\par

As already mentioned in the discussion of figures~\ref{fig4} and
\ref{fig5} the von Schweidler exponent $b'$ and the KWW exponent
$\beta$ are significantly different for the correlators discussed in
these figures. In order to investigate this effect in more detail
we show in Fig.~\ref{fig8} the $q$ dependence of $b'$ and $\beta$ for
the A particles. We clearly recognize that $b'$ is significantly
smaller than $\beta$ for all values of $q$. Thus this is further
evidence that the functional form appropriate to describe the data on
the time scale of the late $\alpha$-relaxation is not appropriate to
describe the data on the time scale of the late
$\beta$-relaxation.\par

Also included in the figure is the KWW exponent $\beta$ for the B
particles. The general behavior of this curve is very similar to the
one for the A particles except that for a given $q$ the values of
$\beta$ for the B particles is a bit smaller that the one for the A
particles. This is in accordance with the prediction of MCT for a
binary mixture of soft spheres \cite{fuchsal93}, namely that the
relaxation of $F_{s}(q,t)$ is more stretched for the small particles
than the one for the large particles.\par

We see in Fig.~\ref{fig8}, that for large values of $q$, $\beta$ for
the A particles is constant to within the noise. We have observed a
similar effect in the case of the $\beta$ determined from $F(q,t)$ for
the AA correlation but {\em not} in the case of $F_{s}(q,t)$ (also
included in Fig.~\ref{fig8}) and $F(q,t)$ for the B particles and BB
correlation, respectively. MCT predicts that for large values of $q$
the KWW exponent should approach as a limiting value
$b$\cite{fuchs94,fuchsal93}. From the figure one recognizes that
unfortunately the quality of our data is insufficient to test this
prediction of the theory.\par

We now investigate the temperature dependence of $\tau$, the relaxation
time of the $\alpha$-relaxation (see Eq.~(\ref{eq7})). In
Ref.~\cite{kob94a} we showed that for low temperatures the relaxation
time of $F_{s}(q,t)$ for the A and B particles for $q=q_{max}$, show a
power-law dependence on temperature, $(T-T_{c})^{-\gamma}$, with a
critical temperature $T_{c}=0.432$, which is very close to the one we
found for the diffusion constant, which was $T_{c}=0.435$. In
Fig.~\ref{fig9} we show the relaxation times versus $T-T_{c}$ for the
following correlation functions: $F_{s}(q,t)$ for the A and B particles
(squares and triangles pointing down, respectively) at $q=q_{max}$ and
$q=q_{min}$ (of the corresponding structure factors), $F(q,t)$ for the
AA and BB correlation (circles and diamonds, respectively) at
$q=q_{max}$ and $q=q_{min}$, and $F(q,t)$ for the AB correlation
(triangles pointing up and star) for $q=q_{min1}$, $q=q_{max}$ and
$q=q_{min2}$. In each case the filled and open symbols correspond
to $q=q_{max}$ and $q=q_{min}$, respectively and the star is for
$q=q_{min2}$. The last three values of $q$ are the location of the
first minimum, the first maximum and the second minimum in $S(q)$ for
the AB correlation (see Fig.~\ref{fig1}b) and have the values 6.40,
7.72 and 12.05. From Fig.~\ref{fig9} we recognize that at low
temperatures the various relaxation times can all be fitted well by a
power-law with the same critical temperature $T_{c}=0.430$. Thus the
critical temperature for all correlators shown in this figure is very
close to the one we found for the diffusion constant of the A and B
particles thus giving evidence that at this temperature the system
becomes nonergodic. This transition from an ergodic to nonergodic
behavior can of course only be expected if the hopping processes will
not restore the ergodic behavior of the system.  We have given evidence
above, that in the temperature interval investigated in this work these
hopping processes are not present.  However, it may very well be that
for temperatures even closer to $T_{c}$ the dynamics of the system is
influenced by the hopping processes and thus no real transition to a
nonergodic behavior will be observed.\par

The critical exponent $\gamma$ with which the various relaxation times
diverge at $T_{c}$, is found to be essentially independent of the type
of correlation function, in accordance with MCT. Its value is around
2.6, which is essentially the same as the exponent predicted by the
theory if one uses the connection between $\gamma$ and the von
Schweidler exponent $b$ and a value of $b$ of 0.51 (leading to
$\gamma=2.7$) which was found for $b$ from $F_{s}(q,t)$ for the A
particles close to $q_{max}$ (see Fig.~\ref{fig4}).\par

We also note that in Ref.~\cite{kob94c} we showed that at low
temperatures the diffusion constant for the A and B particles were
better fitted by a power-law than by a Vogel-Fulcher law. We tried to
fit the relaxation times presented in Fig.~\ref{fig9} also with a
Vogel-Fulcher law and found that for these quantities the Vogel-Fulcher
law was able to make a good fit which covered a slightly larger
temperature range than the power-law presented in Fig.~\ref{fig9} is
able to do.  Thus from the point of view of a mere fitting function the
Vogel-Fulcher law has to be preferred for this data. However, this
should not be taken as an argument for not using a power-law to fit the
data to extract an exponent for comparison with other exponents.  The
objective of the present work is to test whether MCT is able to give a
{\em consistent} picture of the whole variety of data on the low
temperature dynamics of the system.  The fact that one subset of the
data can be better described by a functional form not compatible with
the theory should not be seen as evidence against the validity of the
theory.\par

{}From Fig.~\ref{fig9} we also recognize that the relaxation times for
the various correlation functions depends strongly on the correlator.
In particular we see that at the lowest temperature we find a variation
of about a factor of seven for the $\tau$ of the various correlation
functions presented here. In this context it is interesting to test
another prediction of MCT, the so-called $\alpha$-scale universality
(see Eq.~(\ref{eq7})). In order to do this we defined a new relaxation
time $\tau'$ by requiring that at time $\tau'$ the correlator has
decayed to 0.15. The reason for this definition is
that with the old definition, for which the correlator is supposed to
have decayed to $e^{-1}\approx 0.368$, the relaxation times $\tau$ did
not measure the $\alpha$-relaxation time if $q$ is large, since for
large values of $q$ the height of the plateau becomes smaller than
0.368 (see Fig.~\ref{fig3}).\par

In Fig.~\ref{fig10} we show the $q$ dependence of $\tau'$ determined
from $F(q,t)$ for the AA correlation for all temperatures
investigated. From this figure we recognize that at low temperatures
(top curves) the curves corresponding to different temperatures are
just shifted vertically. This is exactly what is expected if the
$\alpha$-scale universality holds (see Eq.~(\ref{eq7})). Thus at low
temperature the $q$ dependence of each curve is given by $C(q)$ and the
vertical shift comes from the strong temperature dependence of the
factor $(T-T_{c})^{-\gamma}$.\par

In the figure we recognize that at low temperature the individual
curves show quite a few minima and maxima. Some of them can be
identified with the minima and maxima in the structure factor $S(q)$
(see Fig.~\ref{fig1}a). The other extrema can be traced back to the
little oscillations in the structure factor, which we said was a finite
size effect. Thus these other extrema in Fig.~\ref{fig10} should
presumably also be considered to be a finite size effect. It is
remarkable that the $q$ dependence of the relaxation time found here is
qualitatively similar to the one predicted by MCT \cite{fuchsal93}
showing that also this aspect of the theory seems to be, at least
qualitatively, correct. A similar agreement is observed for the
relaxation times of the incoherent intermediate scattering
function.\par

Note that the ratio of the relaxation time at $q=q_{max}=7.25$ and
at $q\approx 16$, the second minimum in $S(q)$, is about 50.  Thus this
shows even more dramatically than it did in Fig.~\ref{fig9}, that the
relaxation times are dependent on $q$. A similar $q$ dependence was
also found in experiments by Mezei {\it et al.} on CKN
\cite{mezei87}.\par

\subsection{Dynamic Susceptibility}
\label{ssec:IV-B}
In this subsection we test the prediction of MCT for the imaginary part
$\chi''(\omega)$ of the dynamic susceptibility $\chi(\omega)$. This
quantity can be obtained by taking the time Fourier transform of the
intermediate scattering function and multiplying it by the frequency
$\omega$. Since in this work the intermediate scattering function
extends, at the lowest temperature, over almost seven decades in time
the computation of the Fourier transform is not a trivial task and
great care has to be taken in order to avoid the generation of spurious
features in $\chi(\omega)$, which in turn might prevent the testing of
certain predictions of the theory. In order to overcome this problem we
parametrized the intermediate scattering function with a spline under
tension \cite{reinsch} and computed the Fourier transform of the latter
by means of the Filon algorithm \cite{abramowitz}.\par

In Fig.~\ref{fig11} we show the dynamic susceptibility
$\chi_{s}''(q,\omega)$ computed from $F_{s}(q,t)$ for the A particles at
$q=q_{max}$ and $q=q_{min}$ and all temperatures investigated. We
recognize that at high temperatures $\chi_{s}''(q,\omega)$ is just a
single peak located at microscopic frequencies. The form of this peak
is approximated very well by a Lorentzian. On lowering the temperatures
this peak starts to split into two peaks. The first one stays at
microscopic frequencies and the second one moves quickly to small
frequencies with decreasing temperature. Thus we observe nicely how the
$\alpha$-peak separates from the microscopic peak. Note that this
splitting into two peaks looks strikingly similar to the result of a
theoretical calculation with a schematic model of MCT, as a comparison
of Fig.~\ref{fig11} with Fig.~2 of Ref. \cite{gotzesjo88} shows. Thus the
theory is able to describe this effect at least qualitatively.  By
comparing Fig.~\ref{fig11}a with Fig.~\ref{fig11}b we recognize that at
high temperatures the height of the microscopic peak does not depend
strongly on $q$. This is not the case at low temperatures where this
height for $q=q_{max}$ is about 30\% smaller than the one for
$q=q_{min}$. On the other hand, just the opposite trend is observed for
the height of the $\alpha$-peak which at low temperatures is
significantly larger for $q=q_{max}$ than for $q=q_{min}$. All these
observations can be easily understood by remembering that the height
$f_{c}$ of the plateau in the intermediate scattering function depends
on $q$.  Since the height of the $\alpha$-peak is proportional to this
height and the height of the microscopic peak is proportional to
$1-f_{c}$ the above described dependence of the heights of the $\alpha$
and microscopic peak follow directly from the $q$ dependence of the
nonergodicity parameter $f_{c}$.\par

{}From Fig.~\ref{fig11} we also recognize that at low temperatures the
shape of the $\alpha$-peak does not depend on temperature. This is the
consequence of the time temperature superposition principle predicted
by MCT, which we showed to hold very well for this system (see
Fig.~\ref{fig4} and \ref{fig5}). In order to investigate this property
more closely we plot in Fig.~\ref{fig12} the dynamic susceptibility for
the A particles for $q=q_{max}$ scaled by its value at the maximum
versus the scaled frequency $\omega/\omega_{max}$, where $\omega_{max}$
is the location of the $\alpha$-peak. From this figure we recognize
that the shape of the peak changes when we go from high to intermediate
temperatures. This change seems to be most pronounced on the high
frequency side of the peak whereas the low frequency side seems to be
essentially independent of temperature in the whole range of
temperatures investigated. For intermediate and low temperatures also
the high frequency side of the peak does not change with temperature.
The only change we observe in the curves when the temperature is
lowered is that the high frequency wing of the peak extends to higher
and higher rescaled frequencies before the curves turn up again to the
microscopic peak. Thus this plot shows that the time temperatures
superposition principle holds very well for this system.\par

It is interesting to compare this figure with results from a
depolarized light-scattering experiment on orthoterphenyl
\cite{steffen94}.  Figure 15 of Ref.~\cite{steffen94} shows how this
type of plot looks like when the time temperature superposition
principle supposedly does not hold. In that figure the curves for different
temperatures fall also onto a master curve on the low frequency side of
the $\alpha$-peak.  However, on the high frequency side of the peak a
clear dependency of the curves on temperature is observed, in contrast
to our findings presented in Fig.~\ref{fig12}.\par

A more quantitative way to show that the time temperature superposition
holds for a given correlator is to measure the full width at half
maximum (FWHM) of the $\alpha$-peak and to investigate the temperature
dependence of this quantity. We define the FWHM as the ratio of the
two frequencies at which a horizontal line at half the height of the
peak intersects the low and high frequency wing of the peak. Note that
the determination of the FWHM at intermediate temperatures is not
always possible as can been seen, e.g., from Fig.~\ref{fig12} since at
these temperatures the microscopic and the $\alpha$-peak are too close
together and thus there is no well separated $\alpha$-peak.  We
determined the FWHM for those eleven correlators for which we also
presented the relaxation times as a function of temperature in
Fig.~\ref{fig9} and show this quantity as a function of temperature in
Fig.~\ref{fig13}. In order to spread the temperature scale at low
temperatures we plot $T-0.435$ on a logarithmic scale.  Note that this
representation has nothing to do with some sort of critical behavior
but is only a convenient way to present the data. The meaning of the
symbols is the same as in Fig.~\ref{fig9}.  More details can be
obtained from the figure caption. We recognize that at high
temperatures the FWHM is relatively small for all correlators. It is
interesting to note that at these temperatures the FWHM of the
correlators for the AB correlation functions (triangles pointing up and
star) are significantly smaller than the ones of the other correlators.
Thus we conclude that at high temperatures the relaxation behavior for
the formers, being non-Debye, is different from the one of the latters,
which are of a Debye type.\par

On lowering the temperature the FWHM increases significantly up to
$T=2.0$. For temperatures between $2.0>T>0.6$ we are not able to measure
the FWHM because of the above mentioned problem. (The only correlator
for which we have data even in this temperature interval has a
microscopic peak which is much smaller that the $\alpha$-peak and does
therefore not interfere significantly with the latter.) For
temperatures $T\leq 0.6$ the FWHM can be considered constant to within
the noise of the data. Thus this is further evidence that the time
temperature superposition principle holds for all correlators
investigated here.\par

MCT predicts that in the time dependence of the function $G(t)$ in
Eq.~(\ref{eq1}) only the ratio $t/t_{\epsilon}$ enters (see
Eq.~(\ref{eq2})), where $t_{\epsilon}$ is the time scale of the
$\beta$-relaxation. Consequently the time Fourier transform of
$G(t/t_{\epsilon})$ will depend only on the ratio
$\omega/\omega_{\epsilon}$. Thus this prediction can be tested by scaling
the dynamic susceptibility by its value at the minimum between the
$\alpha$-peak and the microscopic peak and plotting it versus
$\omega/\omega_{\epsilon}$, where $\omega_{\epsilon}$ is the location of
this minimum. We have done this for $\chi_{s}''(q,\omega)$ for
$q=q_{max}$ for the A particles. The resulting scaling plot is shown in
Fig.~\ref{fig14} for all those temperatures for which a minimum could
be identified, i.e. $T\leq 0.8$. From this figure we recognize that in
the vicinity of the minimum the curves fall indeed onto a master curve.
Thus the scaling behavior predicted by MCT holds for this correlation
function. We made the same type of scaling plot also for the other ten
correlation functions mentioned above and found in all cases that at
low temperatures they fell onto a master curve.\par

Also included in Fig.~\ref{fig14} (dashed line) is a fit with the time
Fourier transform of the theoretical master curve predicted by MCT for
the time Fourier transform of $G(t)$. This theoretical curve depends
again on the exponent parameter $\lambda$ which we fixed to the value
we found for it from the fit of the theoretical master curve in the
time domain, i.e. to $\lambda=0.77$ (see the discussion in the context
of figures \ref{fig4} and \ref{fig5}). This theoretical master curve
was computed by taking into account the first few correction terms to
the so-called interpolation formula \cite{gotze90}. We recognize from
Fig.~\ref{fig14} that for rescaled frequencies to the right of the
minimum the fit with the theoretical master curve is not good at all.
This is another manifestation of the fact stated earlier, that for this
system MCT does not describe well the early part of the
$\beta$-relaxation regime (see the discussion in the context of figures
\ref{fig4} and \ref{fig5}).  A similar discrepancy between the fit of
the spectrum with master curves proposed by MCT and experimental data
was discussed in Ref.~\cite{rossler94} where it was argued that the
reason for this discrepancy was the presence of the so-called Boson
peak.\par

We see from Fig.~\ref{fig14} that
the theoretical master curve does not even give a very satisfactory fit
to the master curve to the {\em left} of the minimum, i.e. where the von
Schweidler law should be observed. The deviation of the theoretical
curve from the master curve of the data should, however, {\em not} be
considered as a flaw of the theory. We have shown in Fig.~\ref{fig4}a
that in the time domain the theoretical curve gives an excellent fit to
the master curve over several decades in time. If we find now that the
corresponding Fourier transforms do not look very similar to each other
this has to be viewed as a unpleasant property of the Fourier
transformation. The problem is that the Fourier transform of the
theoretical master curve was obtained by Fourier transforming a series
expansion (in powers of $t/t_{\epsilon}$) of the theoretical master curve
in the time domain.  The dynamic susceptibility from the simulation,
however, was obtained by taking the Fourier transform of the time
correlation function.  Therefore the dynamic susceptibility computed
in this way will include in the vicinity of the minimum also
contributions from times which are {\em not} in the $\beta$-relaxation
regime. Therefore it cannot be expected that if a theoretical master
curve fits the data well over a certain numbers of decades in time that
then also the corresponding Fourier transform will match over the same
number of decades in frequency. This effect has also be observed in
Ref.~\cite{gotze90} where it was found that the theoretical master
curve approximated the solution of a schematic model \cite{bibles} well
over 5.5 decades in time, but that the corresponding Fourier transforms
matched only over four decades in frequency.\par

We also point out that the above mentioned effect might make the
determination of the exponent parameter $\lambda$ from measurements of
the susceptibility a bit problematic. The difficulty arises from the
fact that one usually tries to obtain a good fit over a frequency
interval that is as large as possible. In doing that, one might
severely overestimate the range in frequency for which the fit with the
theoretical master curve is supposed to hold, which in turn may lead to
a wrong value of the exponent. In order to illustrate this we have
tried to make a fit to the master curve with a theoretical master curve
in which the exponent parameter was a free fit parameter. The result of
this fit is included in Fig.~\ref{fig14} as well (dotted line) and we
recognize that now the high frequency wing of the $\alpha$-peak is
fitted quite well by the theoretical master curve.  The value of
$\lambda$ we obtained was 0.74. At first glance the difference of this
value to the optimal one as determined from the time domain
($\lambda=0.78$) does not seem to be large. However, when we use this
new value of $\lambda$ in order to make a fit in the time domain, the
resulting fit was significantly inferior to the one presented in
Fig.~\ref{fig4}a. Thus this value of $\lambda$ is not compatible with
the data from the time domain. Hence if we had access only to the
susceptibility data, we probably would have determined an incorrect
value of $\lambda$.  It has to be emphasized, however, that the
determination of $\lambda$ from data in the frequency domain might be
much less problematic if the theoretical master curve gives a good fit
on {\em both} sides of the minimum.\par

Apart from the correlator investigated in the context of
Fig.~\ref{fig14} we tested whether the other ten correlation functions
mentioned earlier also showed a master curve when scaled in the
appropriate way, i.e.  by $\chi_{\epsilon}''$ and $\omega_{\epsilon}$, and
found this to be indeed the case.  MCT predicts that these various
master curves should be identical, since they are all related to the
time Fourier transform of the function $G(t)$ (see Eq.~(\ref{eq1})),
which is predicted to be independent of the correlator.  In order to
check this prediction of the theory we show in Fig.~\ref{fig15} the
susceptibilities at $T=0.466$, the lowest temperature investigated,
scaled by its value at the minimum, versus $\omega/\omega_{\epsilon}$.
We recognize from this figure that the overall form of the curves,
e.g.  the height of the $\alpha$-peak or the one of the microscopic
peak, for the various correlators is very different and thus this
prediction of the theory is clearly not a trivial one. We also see that
on the left hand side of the minimum most of the curves follow a master
curve and thus the prediction of the theory is confirmed for these
correlators.  However, there are also a few curves which do not fall
onto this master curve. These curves are found to be the ones for which
the nonergodicity parameter $f_{c}$ is relatively small, and thus the
ones with values of $q$ which are close to the first minimum.  This
part of the master curve, i.e.~left from the minimum, corresponds to
the von Schweidler regime.  Moreover, it is known \cite{buchalla88}
that corrections to the leading asymptotic results of the theory for
the von Schweidler law are large if the nonergodicity parameter is
small.  Thus, the deviation of small-$f_c$ curves from the master curve
can be rationalized within the context of the theory.\par

For rescaled frequencies to the right of the minimum we recognize
that the curves seem to cluster around {\em two} different
``master'' curves. A closer analysis of what type of correlator belongs
to which bunch of curves showed that the lower bunch of curves belong
all to correlators for the incoherent and coherent part of the
intermediate scattering function of the B particles. Since we are not
aware of any predictions of MCT on the size dependence of the
corrections to this master curve we cannot offer any reason for this
behavior and thus only report the observation.\par

Since MCT predicts that at a given temperature the function $G(t)$ is,
for a given system, independent of the correlator, its time Fourier
transform should also be independent of the correlator.  In particular
this means that the location of the minimum in the susceptibility
should be independent of the correlator as well. In order to test this
prediction of the theory we present in Fig.~\ref{fig16} the
susceptibilities for the eleven correlators mentioned above for the
lowest temperature studied in this work, i.e. $T=0.466$. From this
figure we see that the form of the various curves depends strongly on
the type of correlator.  However, to within the noise of the data the
location of the minimum is independent of the type of correlator, thus
confirming this prediction of the theory.\par

The theory predicts that $\chi_{\epsilon}''$, the value of the dynamic
susceptibility at the minimum, should show a  square-root dependence as
a function of temperature (see Eq.~(\ref{eq10})). This prediction can
easily be tested by plotting $(\chi_{\epsilon}'')^{2}$ versus $T$. If the
prediction of MCT is correct one should find straight lines which
intersect the temperature axis at $T=T_{c}$. We made this plot for the
eleven correlators mentioned earlier and found that the curves were not
straight lines, thus contradicting the prediction of the theory. We
also tested whether a different exponent than 2.0 would lead to
straight lines in this type of plot and found that an exponent around
1.25 was indeed able to do so. Interestingly enough the extrapolation of
the resulting straight lines to lower temperatures intersect the
temperature axis all around the temperature $T=0.438$, which is very
close to the critical temperature we found for the diffusion
constant~\cite{kob94c,kob94a} or the relaxation times (see
Fig.~\ref{fig9}), which were 0.435 and 0.430, respectively. This result
is presented in Fig.~\ref{fig17}, were we plot the $\chi_{\epsilon}''(T)$
versus $T-T_{c}$ in a double logarithmic way for all the eleven
correlators investigated. The solid line is a power-law with exponent
0.8=1/1.25, showing that the slope of the various curves is close to this
value. (If the prediction of MCT were correct this slope should be
0.5).\par

The theory also predicts that $\omega_{\epsilon}$ and $\chi_{\epsilon}''$
are connected via a power-law and that the exponent is the critical
exponent $a$ from MCT (see Eq.~(\ref{eq12})). In Fig.~\ref{fig18} we
show that $\chi_{\epsilon}''$ and $\omega_{\epsilon}$ are indeed related by
a power-law but that the exponent is close to unity (solid line) and
thus much larger than the one predicted by MCT for this system (which,
with $b\approx 0.49$, would be around 0.28). Note that an exponent of
1.0 is expected if we assume that the power-law seen to the right of
the minimum is not the one predicted by MCT but just the low frequency
wing of a microscopic peak that is the Fourier transform of a process
whose integral in the time domain is finite. One possibility of such
a process is thus a Debye-like process. Since we have already
seen earlier (see the discussion in the context of figures~\ref{fig4}
and \ref{fig5}) that it may well be that the critical decay predicted
by MCT is, at the temperatures investigated here, still severely
disturbed by the Debye-like relaxation behavior occurring at short
times, this explanation is in accordance with our previous finding.\par

A further test of the theory is to investigate the connection between
the frequency $\omega_{\epsilon}$ and $\omega_{max}$, the frequency of
the $\alpha$-peak. MCT predicts this connection to be also a power-law
but this time with an exponent $b/(a+b)$ which, assuming $b\approx
0.49$, is around 0.64 (see Eq.~(\ref{eq13})). Fig.~\ref{fig19} shows
that for our system the power-law is indeed observable but that the
exponent is around 0.33 (solid line). This value can again be
understood by assuming that the value of $a$ is 1.0, or in other words
that the high frequency wing of the minimum is just the low frequency
part of the Debye-like microscopic peak thus giving an exponent of
0.33.\par

To conclude we test whether the prediction of the theory that
$\omega_{max}$ should show a power-law dependence on temperature with
an exponent $\gamma$ (see Eq.~(\ref{eq11})) holds for this system. We
tried to fit the low temperature behavior of $\omega_{max}$ with such a
functional form and the result is presented in Fig.~\ref{fig20}.
{}From this figure we recognize that the power-law predicted by the
theory indeed holds and that also the exponent is in accordance with
the one which follows from the prediction of MCT with the von
Schweidler exponent $b\approx 0.49$, giving $\gamma=2.7$. The critical
temperature is the same as the one found for the constant of diffusion
\cite{kob94c,kob94a} and very close to the one found for the relaxation
times and the one for $\chi_{\epsilon}$, which is in accordance with the
theory.

\section{Summary and Conclusions}
\label{sec:V}
We have presented the results of a large scale molecular dynamics
computer simulation of a supercooled binary Lennard-Jones mixture. The
goal of our investigation was to test whether MCT is able to give a
correct description of the dynamics of this simple liquid at low
temperatures. In contrast to an earlier paper \cite{kob94c} in which we
mainly concentrated on the investigation of the diffusion constant
and the van Hove correlation function, we focus in this work on the
intermediate scattering function and the dynamic susceptibility. This
allows us to perform additional tests in order to investigate whether
MCT is able to describe the low-temperature dynamics of our system.
In this section we summarize the salient results of our
investigation.\par

We find that at low temperatures $F_{s}(q,t)$ and $F(q,t)$, the
incoherent and coherent intermediate scattering function, respectively,
show the two step relaxation process predicted by MCT.  By scaling the
intermediate scattering function by the $\alpha$-relaxation time
$\tau(q,T)$ we find that the correlators for intermediate and low
temperatures fall onto a master curve in the $\alpha$-relaxation
regime. Thus the time temperature superposition principle predicted by
MCT holds for this system. As predicted by the theory the early part of
this master curve, which in the language of MCT corresponds also to the
late $\beta$-relaxation regime, is fitted very well by a power-law, the
so-called von Schweidler law. Also the functional form predicted by MCT
for the master curve in the $\beta$-relaxation regime, the so-called
$\beta$-correlator, which takes into account the corrections to the von
Schweidler law, gives a very good fit to the master curve in the region
of the late $\beta$-relaxation. From the point of view of the quality
of the fit, the two functional forms can be considered as equally good.
Thus for those correlators investigated the corrections to the von
Schweidler law do not seem to be very important for this system.
Computing the von Schweidler exponent $b$ from the exponent parameter
$\lambda$ and comparing it to the exponent $b'$ as determined from the
power-law fit we find that $b$ and $b'$ are very close together for
those correlators and values of $q$ for which we made both kinds of
fits. Thus we can take $b'$ as a substitute for $b$ and, because of the
connection between $b$ and $\lambda$ (see Eq.~(\ref{eq5})), investigate
the dependence of $\lambda$ on $q$ and the type of correlator by the
investigation of the dependence of $b'$ on these quantities.  For
values of $q$ in the range of $q_{max}$ and $q_{min}$, the location of
the first maximum and the first minimum of the structure factor,
respectively, the exponent $b'$, shows only a weak dependence on $q$ or
the type of correlator \cite{kob94a}.  If $q$ is varied over
a larger range, however, one finds that $b'$ depends on $q$, which is in
contradiction with the prediction of MCT.  On the other hand it has to
be remembered that for $q$ very small and $q$ very large the theory
predicts the existence of correction terms to the $\beta$-correlator
and thus these apparent deviations might just be the result of these
correction terms.\par

For the early part of the $\beta$-relaxation regime the theory predicts that
the correlators should show a different power-law, the so-called
critical decay (see Eq.~(\ref{eq3})). We do not find any hint that, in
the temperature range studied, our system shows such a time
dependence.  Rather we find that the fast Debye-like relaxation behavior at
short
times goes directly over to the slow decay of the late
$\beta$-relaxation regime. Thus it might be that for our system the critical
decay is just not existent at all or that, in the temperature range we
are able to investigate, the critical decay is not visible because of
the interference with the dynamics at short times.  It is interesting
to note that the numerical solution of the mode-coupling equations in
which the full $q$ dependence and a reasonable short time behavior were
taken into account showed that the observation of the critical decay is
very difficult if the thermodynamic state of the system is not very
close to the critical point of MCT \cite{bengtzelius86,smolej93}.  Thus
the fact that we are not able to see the critical decay should not
necessarily be seen as a failure of the theory.\par

It is interesting to note that the critical decay seems to be more
readily observable if the dynamics of the particles is not Newtonian
but stochastic. This has been observed, e.g., in a simulation by
L\"owen {\it et al} of a colloidal suspension in which the two
different types of dynamics were compared~\cite{lowen91}. It was found
that the stochastic dynamics led to a relaxation behavior for which the
approach to the plateau is much slower than the one for the Newtonian
dynamics.  Also a recent simulation by Baschnagel of a polymer system
with stochastic dynamics showed a very slow approach to the plateau and
it was demonstrated that in this time region the correlators could be
fitted well with a power-law \cite{baschnagel94}. Very recently G\"otze
and Sj\"ogren were able to show within the framework of MCT that a
stochastic dynamics will lead to a relaxation behavior of the
correlators for which the critical decay is more easily observable than
for a Newtonian dynamics, thus offering a theoretical explanation for
these observations \cite{goetze95}.\par

The late part of the $\alpha$-relaxation regime can be fitted very well
with a KWW-law, in accordance with MCT. The exponent $\beta$ of the
KWW-law is significantly different from $b$, the exponent of the von
Schweidler law. Thus we conclude that, in accordance with the theory,
the von Schweidler law is not the short time expansion of the KWW-law.\par

The height of the plateau, i.e. the nonergodicity parameter $f_{c}$, is
strongly dependent on the wave-vector $q$. As predicted by the theory
the nonergodicity parameter of $F_{s}(q,t)$, often also called the
Lamb-M\"ossbauer factor, shows a Gaussian-like decay in $q$. Also the
$q$-dependence of the nonergodicity parameter for $F(q,t)$, the
so-called Debye-Waller factor, is in qualitative accordance with the
prediction of the theory, in that it shows an oscillatory behavior which
is in phase with the structure factor.\par

The $\alpha$-relaxation time $\tau(T)$ shows for all correlators
investigated at low temperatures a power-law dependence on $T$, as
predicted by the theory. The critical temperature, as well as the
critical exponent $\gamma$, are independent of the correlator. The critical
temperature is very close to the one we determined for the diffusion
constant for both types of particles \cite{kob94c,kob94a}, which is
also in accordance with the prediction of the theory. The critical
exponent $\gamma$ of the power-law of $\tau(T)$ and the von Schweidler
exponent $b$ fulfill the connection put forward by MCT between the two
quantities (see Eq.~(\ref{eq8})), provided that one uses as a value of
$b$ the ones found in the vicinity of $q_{max}$ and $q_{min}$.
However, the exponent is not the same as the one we found for the
critical behavior of the diffusion constant \cite{kob94c,kob94a},
which is in conflict with the prediction of the theory.\par

By investigating the $q$ dependence of the relaxation time $\tau$, we
find that this quantity has a strong dependence on $q$. We also
demonstrate that $\tau$ obeys the $\alpha$-scale universality (see
Eq.~(\ref{eq7})), as predicted by the theory.\par

We show that the time temperature superposition principle can also be
seen very well in the dynamic susceptibility $\chi''(q,\omega)$, in that
we show that at low temperatures the width of the $\alpha$-peak does
not depend on $T$ to within the noise of the data.\par

By scaling frequency by $\omega_{\epsilon}$, the location of the minimum
in $\chi''(q,\omega)$, and scaling $\chi''(\omega)$ by
$\chi''_{\epsilon}(q)=\chi''(q,\omega_{\epsilon})$, we show that at low
temperatures the curves for different temperatures fall onto a master
curve. Thus we demonstrate that the scaling behavior predicted by MCT
holds for our model. A fit with the functional form of MCT is not able
to give a very good fit to this master curve when the exponent
parameter $\lambda$ is fixed to the value we determined from fits to
the correlators in the time domain. This discrepancy is, however, not a
flaw of the theory but traced back to an unpleasant property of the
Fourier transformation of a time correlation function.  If the
parameter $\lambda$ is allowed to float, we are able to generate a
satisfactory fit in the frequency domain, but with the result, that the
$\lambda$ determined in this way is not optimal anymore in the time
domain. Thus we conclude that if the theoretical curve does not give a
good fit to the high frequency side of the minimum, but only gives a
good fit on the low frequency side of the minimum, the determination of
$\lambda$ from fits to the master curve in the frequency domain is
problematic in the sense that it might yield a wrong value of
$\lambda$.\par

In contrast to the prediction of MCT $\chi_{\epsilon}''$ does not show a
power-law dependence on temperature with an exponent $0.5$ but rather
with an exponent 0.8. However, the critical temperature is, as
predicted by the theory, very close to the critical temperature for
the diffusion constant or the relaxation time.\par

To summarize we can say, that MCT is able to describe the dynamics of
our system at low temperatures in a surprisingly accurate way. There
seems to be some differences between the behavior of our system and the
predictions of the theory. However, all these discrepancies can be
rationalized by taking into account that for values of $q$
significantly different from $q_{max}$ and $q_{min}$ there are
important corrections to the asymptotic results of the theory and that
the theory is valid only very close to $T_{c}$. Understanding whether
the observed discrepancies can really be understood within the
framework of the theory or whether the theory has reached its limit of
applicability is clearly of great interest, and we hope that this
question can be answered in the future. One possible way to address
this question is to solve numerically the mode coupling equations,
in which the full $q$-dependence is taken into account, and compare these
solutions with the results of our simulation. This work is currently
in progress \cite{nauroth95}.

\acknowledgements
We thank Dr. J. Baschnagel, Dr. M. Fuchs and Prof. W. G\"otze for many
useful discussions and a critical reading of the manuscript.  Part of
this work was supported by National Science Foundation grant
CHE89-18841. We made use of computer resources provided under NSF grant
CHE88-21737.

\begin{figure}
\noindent
\caption{Structure factor $S(q)$ for all temperatures investigated. For
clarity the individual curves have been displaced vertically by $x\cdot
n$ with $n=0,1,2,\ldots$. a) AA correlation, $x=0.2$; b) AB correlation,
$x=0.1$; c) BB correlation, $x=0.025$.\label{fig1}}
\vspace*{5mm}
\par
\caption{Incoherent part of the intermediate scattering function
$F_{s}(q,t)$ for all temperatures investigated. a) and b): A particles,
$q=q_{max}=7.25$ and $q=q_{min}=9.61$, respectively. c) and d): B particles,
$q=q_{max}=5.75$ and $q=q_{min}=7.06$, respectively. \label{fig2}}
\vspace*{5mm}
\par
\caption{Incoherent intermediate scattering function $F_{s}(q,t)$ for
the A particles at $T=0.466$. The values of $q$ range from $q=6.0$
(top) to $q=14.0$ (bottom) and are given by $q=6.0+0.36n$ with
$n=0,1,2\ldots$.\label{fig3}}
\vspace*{5mm}
\par
\caption{Incoherent intermediate scattering function $F_{s}(q,t)$ for
all temperatures investigated (solid lines) versus rescaled time. The
dashed curve is a fit with a master curve in the $\beta$-relaxation
regime proposed by MCT (see text for details). The dotted curve is a
fit with a von Schweidler law and the chained curve is a fit with a KWW
law. a) and b): A particles, $q=q_{max}=7.25$ and $q=q_{min}=9.61$,
respectively. c) and d): B particles, $q=q_{max}=5.75$ and
$q=q_{min}=7.06$, respectively.  \label{fig4}}
\vspace*{5mm}
\par
\caption{Coherent intermediate scattering function $F(q,t)$ for all
temperatures investigated (solid lines) versus rescaled time.  The
dashed curve is a fit with a master curve in the $\beta$-relaxation
regime proposed by MCT (see text for details). The dotted curve is a
fit with a von Schweidler law and the chained curve is a fit with a KWW
law. a) AA correlation for $q=q_{max}=7.25$, b) AB correlation for
$q=q_{max}=7.62$ and c) BB correlation for
$q=q_{max}=5.75$.\label{fig5}}
\vspace*{5mm}
\par
\caption{Nonergodicity parameter $f_{c}$ for the incoherent (A particles)
and coherent (AA correlation) intermediate scattering
function (upper solid line and upper dotted line, respectively).
The lower solid line and lower dotted line are the amplitudes of a
KWW fit at long times to the same incoherent and coherent
intermediate scattering function. The dashed line is the structure
factor $S(q)$ for the AA correlation divided by 2.0.\label{fig6}}
\vspace*{5mm}
\par
\caption{Effective von Schweidler exponent $b'$, as determined from a
power-law fit, versus $q$ for the incoherent intermediate scattering
function for the A and B particles (curve A and B, respectively) and
for the coherent intermediate scattering function for the AA, AB and BB
correlation (curves AA, AB and BB).\label{fig7}}
\vspace*{5mm}
\par
\caption{KWW exponent $\beta$ determined from the incoherent
intermediate scattering function for the A and B particles and the
effective von Schweidler exponent $b'$ for the same correlator for the
A particles. \label{fig8}}
\vspace*{5mm}
\par
\caption{Relaxation time $\tau$ versus temperature for various
correlators. Squares and triangles pointing downwards:  $F_{s}(q,t)$
for A and B particles, respectively. Circles and diamonds: $F(q,t)$ for
AA and BB correlation, respectively.  Triangles pointing upwards and
star: $F(q,t)$ for AB correlation.  Filled and open symbols are for
$q=q_{max}$ and $q=q_{min}$, respectively.  Solid line: power-law with
exponent 2.6.\label{fig9}} \vspace*{5mm}
\par
\caption{Relaxation time $\tau'$ (see text for its definition) for the
coherent intermediate scattering function for the AA correlation versus
$q$ for all temperatures investigated.\label{fig10}}
\vspace*{5mm}
\par
\caption{Dynamic susceptibility $\chi_{s}''(q,\omega)$ for the A
particles versus $\omega$ and all temperatures investigated. a)
$q=q_{max}$ b) $q=q_{min}$. \label{fig11}}
\vspace*{5mm}
\par
\caption{Dynamic susceptibility $\chi_{s}''(q,\omega)$ scaled by
$\chi_{max}''$ for $q=q_{max}$ for the A particles versus rescaled
frequency $\omega/\omega_{max}$ for all temperatures
investigated.\label{fig12}}
\vspace*{5mm}
\par
\caption{Full width at half maximum of the $\alpha$-peak for various
dynamic susceptibilities (see text). The meaning of the symbols is
the same as in Fig.~9.\label{fig13}}
\vspace*{5mm}
\par
\caption{Solid lines: dynamic susceptibility $\chi_{s}''(\omega)$ for
$q=q_{max}$ for the A particles scaled by its value at the minimum
versus frequency scaled by the location of this minimum. Dashed line:
Fit with theoretical master curve from MCT with exponent parameter
$\lambda$ fixed to the value determined from the corresponding fit in
the time domain (see Fig.~4a). Dotted curve: Fit with the theoretical
master curve from MCT with $\lambda$ as free fit
parameter.\label{fig14}}
\vspace*{5mm}
\par
\caption{Dynamic susceptibility $\chi''(q,\omega)$ for the eleven
correlators investigated (see Fig.~9 for details) scaled by its value
at the minimum versus frequency scaled by the location of this
minimum.\label{fig15}}
\vspace*{5mm}
\par
\caption{Dynamic susceptibility $\chi''(q,\omega)$ for various
correlators (see text for details) at the lowest temperature
$T=0.466$.\label{fig16}}
\vspace*{5mm}
\par
\caption{$\chi_{\epsilon}''$, the value of the dynamic susceptibility at
the minimum, versus $T-T_{c}$ for eleven correlators (see Fig.~9 for
details) in a double logarithmic plot. Solid line: power-law with
exponent 0.8.\label{fig17}}
\vspace*{5mm}
\par
\caption{$\chi_{\epsilon}''$, the value of the dynamic susceptibility
at the minimum, versus $\omega_{\epsilon}$, the location of this
minimum for eleven correlators investigated (see Fig.~9 for details).
Solid line: power-law with exponenet 1.0.\label{fig18}}
\vspace*{5mm}
\par
\caption{$\omega_{\epsilon}$, the location of the minimum in the
dynamic susceptibility, versus $\omega_{max}$, the location of the
$\alpha$-peak, for eleven correlators investigated (see Fig.~9 for
details). Solid line: power-law with exponent 0.33.\label{fig19}}
\vspace*{5mm}
\par
\caption{Temperature dependence of $\omega_{max}$, the location of the
$\alpha$-peak, for eleven correlators investigated (see Fig.~9 for
details). Solid line: power-law with exponent 2.5.\label{fig20}}
\end{figure}

\begin{references}
\bibitem{footnote1}
Note that our definition of the van Hove correlation function is
slightly different from the one usually found in the literature
(e.g. Ref.~\cite{hansenmcdonald86}). See Ref.~\cite{kob94c} for details.
\bibitem{footnote2}
These theoretical curves were computed with a program written by M.
Fuchs and provided to us by the group in Munich. An alternative
method would be to use the tables of Ref.~\cite{gotze90}.
\end{references}
\end{document}